\documentclass[journal abbreviation]{copernicus}

\begin{document}

\title{Thermodynamic origin of life}
\author{K.~Michaelian}
\maketitle

\begin{abstract}
Understanding the thermodynamic function of life may shed light on its
origin. Life, as are all irreversible processes, is contingent on entropy
production. Entropy production is a measure of the rate of the tendency of
Nature to explore available microstates. The most important irreversible
process generating entropy in the biosphere and, thus, facilitating this
exploration, is the absorption and transformation of sunlight into heat.
Here we hypothesize that life began, and persists today, as a catalyst for
the absorption and dissipation of sunlight on the surface of Archean seas.
The resulting heat could then be efficiently harvested by other irreversible
processes such as the water cycle, hurricanes, and ocean and wind currents.
RNA and DNA are the most efficient of all known molecules for absorbing the
intense ultraviolet light that penetrated the dense early atmosphere and are
remarkably rapid in transforming this light into heat in the presence of
liquid water. From this perspective, the origin and evolution of life,
inseparable from water and the water cycle, can be understood as resulting
from the natural thermodynamic imperative of increasing the entropy
production of the Earth in its interaction with its solar environment. A
mechanism is proposed for the reproduction of RNA and DNA without the need
for enzymes, promoted instead through UV light dissipation and diurnal temperature 
fluctuations of Archean sea-surface.
\end{abstract}

\sloppy

\affil{Instituto de F\'{\i}sica, Universidad Nacional Aut\'{o}noma de
M\'{e}xico, Cto.~de la Investigaci\'{o}n Cient\'{\i}fica, Cuidad
Universitaria, Mexico}

\correspondence{K.~Michaelian (karo@fisica.unam.mx)}

\runningtitle{Thermodynamic origin of life} \runningauthor{K.~Michaelian}



\introduction

Only 27 years after the
publication of ``On the Origin of Species'' (Darwin, 1859), Boltzmann
(1886) recognized that the struggle for existence was not a struggle for raw
material, neither for energy, but rather a struggle for entropy (low entropy) 
which became available through the dissipation of high energy
photons to low energy ones (entropy production) through irreversible 
processes occurring within the biosphere. It is now well established that non-equilibrium 
structuring of matter in space and time -- from molecules to hurricanes to living systems
-- is contingent on entropy production (Prigogine, 1967; Prigogine et al., 1972). Such irreversible processes 
are known as ``dissipative structures'' since, although they exist at low entropy, they arise
spontaneously to provide new pathways to a greater sampling of the enormous
multitude of microstates that underlie Nature and, thus, their formation 
increases the entropy of the Universe. 

Onsager (1931) has shown how diverse irreversible processes can couple in order to 
remove impediments to greater global entropy production (Morel and Fleck, 1989). 
In general, the more complex the dissipative structuring in
space and time (i.e.~involving many coupled irreversible processes with 
embedded hierarchal levels and interactions
of long spatial and temporal extent) the greater the overall
entropy production in the systems interaction with its external environment
(Onsager, 1931; Prigogine et al., 1972; Lloyd and Pagels, 1988).

Empirical evidence from the fossil record of the evolutionary history of
Earth indeed suggests that living systems, from cells to the biosphere, have
generally increased in complexity over time, and correspondingly, there has
been an increase in their total entropy production, as well as in the net
entropy production per unit biomass (Zotin, 1984). Onsager's principle, and 
variations thereof, have been useful in describing the existence and 
stability of abiotic and biotic dissipative systems on Earth (Lorenz, 1960; 
Paltridge, 1979; Ulanowicz and Hannon, 1987; Swenson, 1989; Kleidon and 
Lorenz, 2005; Michaelian, 2005; Martyusheva and Seleznev, 2006; Kleidon, 2009).

The ubiquity of the  empirical evidence suggesting that Nature 
tends to find new abiotic, biotic, and coupled abiotic-biotic
pathways to entropy production, is taken here as sufficient justification
for the proposition that RNA and DNA, and their coupling
to the water cycle, arose originally as structuring of material in space and
time to provide a new route to augmenting the entropy production of the
Earth in its interaction with its prebiotic solar environment.

The nucleic acids RNA and DNA
are efficient absorbers of photons in the 200--300\,nm ultraviolet region of
the Sun's spectrum (Chang, 2000); just that important part of the 
spectrum which could have filtered through the early
Earth's dense atmosphere. These molecules, in the presence of water, are
extraordinarily rapid at dissipating the high energy photons to heat. It is then
plausible that life arose as a catalyst for absorbing sunlight at the surface
of the shallow seas, dissipating it into heat and, thereby promoting still
other irreversible processes such as the water cycle (evaporation/rain), and
wind and ocean currents, all of which contribute to the entropy production
of the biosphere (Peixoto et al., 1991; Kleidon 2009). This suggests a
thermodynamic imperative for an origin of life which can be related to its
thermodynamic function of entropy production (Michaelian, 2009).

Prevailing scenarios for the origin of life center on the replication first ``RNA World'' 
(Gilbert, 1986) hypothesis (see Orgel (2004) and Rauchfuss (2008) for a 
review), and ``metabolism first'' followed by replication proposals (see Shapiro (2007)
for a review) including hydrothermal vent proposals (W\"{a}chtersh\"{a}user, 2006;
 LaRowe and Regnier, 2008; Nitschke and Russell, 2009).
The scenario presented here is more akin to the RNA World hypothesis but shares
with the metabolism first theories the idea of a gradual genetic take-over of a metabolic
process. However, replication and metabolism are linked from
the very beginning; the first molecules of life, RNA and DNA, are suggested to be
UV light metabolizing photoautotrophs. 

From within the 
framework of the ``RNA World'', Orgel (2004) recognizes several severe problems
related to low yields in the individual steps of RNA synthesis and replication but is
cautiously optimistic with regard to the abiogenic synthesis of RNA,
suggesting that other, undiscovered, routes to these molecules may
eventually be found. The most difficult current problems with abiogenisis
are; 1)~the production and stability of
ribose over the other more easily synthesized and more stable sugars, 2)~the
difficulty of the polymerization of nucleotides leading to polynucleotides,
3)~the problem of the racemic mixture of chiral nucleotides frustrating the
template-directed copying of polynucleotides and, perhaps the most
difficult, 4)~the replication of RNA without the assistance of enzymes.

The theory presented here offers a consistent framework within which each
of the above mentioned difficulties is alleviated. It recognizes a non-equilibrium
thermodynamic imperative for producing RNA aand DNA, irrespective of the
difficulty of producing these in the laboratory, due to 
the great entropy producing potential of these molecules
given the particular boundary and initial
conditions of the primitive Earth; a high flux of UV photons
and a high sea-surface temperature. Considering these
thermodynamic forces not only alleviates the difficulty with the
abiogenic yields of the primary molecules of life but, at the same time,
provides an UltraViolet and Temperature Assisted mechanism for Reproduction
(UVTAR) of RNA and DNA, without the need for enzymes.

\section{Ambient conditions of insipient life}

Since life obtains its vitality only in the context of its interaction with
its external environment, establishing the ambient conditions of the
primitive Earth is essential to any theory on the origin of life. It has
been hypothesized that the early Earth's seas, existing at the very
beginnings of life some 3.8\thinspace billion years ago (Schidlowski et al.,
1983; Schidlowski, 1988), were hot (Knauth and Lowe, 2003) soups of organic
material.  Inspired by Oparin's (1924) materialistic ideas on
the origin of life, Miller and Urey (1959) experimentally tested and
confirmed the idea that the organic molecules could have been created by
lightning strikes and photochemical reactions on a reducing Archean
atmosphere containing much hydrogen in the form of ammonia (NH$_{3})$ and
methane (CH$_{4})$. Or\'{o} (1961) and Or\'{o} and Kimball (1962) have
demonstrated that all the nucleic acid bases can be obtained by mixing
hydrogen cyanide (HCN) with cyanogen (C$_{2}$ N$_{2}$) and cyanoacetylene (HC%
$_{3}$ N) in an aqueous solution (see also Matthews, 2004). These precursor
cyano-molecules are common products of a reducing atmosphere subjected to UV
photons (Stribling and Miller, 1987; Orgel, 1994).

An alternative hypothesis is that organic molecules are formed in
circumstellar envelopes of stars which shed much of their early 
atmospheres into interstellar space (Hoyle and Wickramasinghe,
1978; Kwok, 2004). In corroboration, it has been found
that regions of high organic molecular densities are often associated with
regions of stellar formation (Ehrenfreund and Charnley, 2000). Over 140
organic molecules have now been found in space, although the nucleobases
have yet to be detected (Kwok, 2009). The dispersed organic molecules are
then suggested to have been deposited into the Earth's oceans through the vehicle
of colliding comets, asteroids, or space dust. Amino acids and nucleobases have been
found in carbonaceous chondrite meteorites, believed to be derived from comets,
such as the Murchinson meteorite (Martins et al., 2008).

Both theories for the origin of the original organic materials are thus
viable and supported by empirical evidence, but which of the two theories is
the most plausible in rendering the high concentration of organic molecules
required in the original soup depends on how reducing the atmosphere was at
the beginnings of life, a topic still highly debated with new insights often
changing the balance (Sagan, 1997; Tian et al., 2005; Cleaves et al., 2008).
For example, recent analysis by Tian et al.~(2005) indicates that the
Earth's early atmosphere may have been colder than originally assumed and,
thus, could have retained up to 30{\%} hydrogen by mass. In this case, the
most likely scenario would have been the production of organic molecules
through lightning and photochemical reactions on H$_{2}$ and CO$_{2}$ rather
than on the ammonia and methane atmosphere assumed in the original Miller
experiments (Tian et al., 2005). Cleaves et al. (2008) have shown that even
in a very neutral atmosphere, the abiogenic yields of the amino acids
can be increased by up to two orders of magnitude if oxidation inhibitors,
such as ferrous iron, are present.

It is generally believed that  Earth's early atmosphere was perhaps
twice as dense as today (Walker, 1977, 1983, 1985), and
containing a similar amount of nitrogen as today, but significantly greater
amounts of CO$_{2}$ and water vapor, perhaps some ammonia and methane, and
probably hydrogen (Cnossen et al., 2007; Haqq-Misra et al., 2008).

Both the composition and density of the early Earth's atmosphere
have relevance to the spectrum of sunlight that could have penetrated to the
Earth's surface. CO$_{2}$~has~a very large photon extinction coefficient at
wavelengths shorter than approximately 202\,nm (at one atmosphere and
295\,K). For longer wavelengths it is found that the optical extinction
demonstrates a $1/\lambda ^4$-like behavior typical for Rayleigh scattering
(Ityaksov et al., 2008). Water vapor absorbs strongly in the ultraviolet
below approximately 170\,nm, and strongly in the infrared above about
1000\,nm, but is basically transparent between these limits
(Chaplin, 2009). Ammonia, NH$_{3}$, also absorbs strongly below 200\,nm.

The early Earth was probably much more volcanically
active than it is today due to the internal heat of accretion, asteroid
bombardment and a higher internal radioactivity. The most important
components of volcanic out-gassing are carbon dioxide CO$_{2}$, sulfur
dioxide SO$_{2}$, and water vapor H$_{2}$O. Sulfur dioxide has a photon
absorption cross section that is large at wavelengths less than
225\thinspace nm with a smaller absorption peak at 290\thinspace nm at
1\thinspace atmosphere and 295\thinspace K (Rufus et al., 2003). As in the
case of present day Venus, photochemical reactions on carbon dioxide, sulfur
dioxide and water vapor can produce sulfuric acid which condenses in a cold
upper atmosphere to produce a fine aerosol that is highly reflective in the
visible region of the Sun's spectrum. Venus has a thin layer of sulfuric
acid cloud at a height of around 70\thinspace km giving the planet an albedo
of 0.77 in the visible and leaving the day-time surface of the planet in
considerable darkness (similar to a very dark overcast day on Earth).
However, the sulfuric acid clouds do not scatter as strongly in the
ultraviolet as in the visible (Shimizu, 1977), a fact made evident by the
observation of Venus in the ultraviolet by Franck~E.~Ross in the 1920's,
which revealed a structure in the cloud cover of the planet for the first
time. Such a layer of reflective sulfuric acid clouds probably existed on
the early Earth. It is known, for example, that a single strong volcanic
eruption on Earth can decrease the global temperature by as much as
0.7\thinspace $^{\circ }$C for several years due to an increase in albedo
in the visible region (Stothers, 1984). 

Water vapor and dense clouds of water on the hot
early Earth, although reflecting uniformly over all wavelengths (hence the 
whiteness of clouds) would have absorbed preferentially in the infrared 
region of the solar spectrum.

The early Sun was more active due to a higher rotation rate and its spectrum
was probably more intense than it is now in the ultraviolet
(Tehrany et al., 2002) and up to 25--30{\%} less intense in the visible (Sagan  and Chyba, 1997). A
larger magnetic field, due to a higher rotation rate, would mean that gamma
and X-ray bursts would also have been much more prevalent and, through
degradation in the Earth's atmosphere, would have lead to an additional
important component of ultraviolet light on the Earth's surface. 
A near-by massive star producing up to $10^{10}$ the amount of UV light as the Sun
could also have been an important contributor to the UV flux on Earth (Zahnle et al., 2007).  

With the water vapor and clouds absorbing strongly in the infrared and the clouds of
sulfuric acid reflecting strongly in the visible, and carbon
dioxide, water, and ammonia absorbing strongly in the ultraviolet below about
200\,nm and sulfur dioxide absorbing below 225\,nm, it is probable that an
enthropically important part of the Sun's spectrum reaching the surface of
the Archean Earth was that in the ultraviolet between approximately 200\,nm
to 300\,nm.

By considering the probable existence of hydrogen sulfide, 
being the thermodynamically stable sulfur-containing gas under
reducing conditions, and the probable formation of aldehydes (formaldehyde
and acetaldehyde) through UV photochemical production, 
Sagan (1973) calculated a somewhat reduced window of
transparency for UV light of between 240 and 290\,nm.

Cnossen et al.~(2007) have carried out detailed simulations of photon
absorption and scattering for various hypothetical models of the Earth's
early atmosphere. Their models consider different CO$_{2}$ concentrations at
different pressures and include absorption, Rayleigh scattering, and an
estimate of the effects of multiple scattering, besides taking into account
the best estimates for the increase in UV intensity expected for a young
Sun. Their conclusions are that the Earth's surface during the Archean %
\mbox{(4--3.5\,Ga)} was subjected to ultraviolet radiation within the
200\thinspace nm to 300\thinspace nm region of up to 10$^{31}$ times 
(at 255 nm) that of present. This is not surprising since today less
than one part in 10$^{30}$ of the incident solar radiation at 250\thinspace
nm penetrates the Earth's ozone and O$_{2}$ atmosphere (Chang, 2000).

The question has arisen as to whether this intense ultraviolet light
had a detrimental effect; through
photolysing, prejudicial photochemical reactions, and too large
informational mutation rates (Biondi et al., 2007), or beneficial effect; through
promoting necessary photochemical reactions, such as abiogenic synthesis of
the nucleic acid bases, ribose and other carbohydrates (Schwartz, 1995), as
well as a favorable selective pressure (Sagan, 1973), on the first molecules
of life (Cockell, 1998; Cnossen et al., 2007; Mart\'{i}n et al., 2009). It
is argued below, that, apart from inducing useful
photochemical reactions and providing favorable selective pressure,
ultraviolet light was crucial to the origin of life for another reason; RNA
and DNA are unparalleled ultraviolet light absorbing molecules (peak absorption at
260 nm) which, in the
presence of water, convert this light rapidly into heat (Middleton et al.,
2009) thereby promoting evaporation. The replication
of RNA and DNA on the sea-surface would thus have been thermodynamically favored
in the intense UV light environment of early Earth since, by coupling to the water
cycle, it supplied a new route to greater entropy production (Michaelian, 2009).

\section{The sea-surface and entropy production}

The sea-surface skin layer, of a thickness of 1\thinspace mm, being
the region of mass, energy and momentum transfer with the atmosphere 
(Hardy, 1982; Soloviev and Lukas, 2006), is of
particular importance to the theory presented here. The upper 50 $\mu$m 
of this layer (the microlayer) hosts an ecosystem of a particularly high organic 
density, up to 10$^{4}$ the density of water only slightly below (Hardy, 1982;
Grammatika and Zimmerman, 2001). The organic material consists of 
cyanobacteria, diatoms, viruses, free floating RNA and DNA and other
living and non-living organic material such as lipids, aldehydes,
chlorophyll and other pigments. A high enrichment of trace metals is also 
found in the microlayer (Hardy, 1982). The high density of material at
the surface is attributed to natural buoyancy and surface tension, but most
notably to the scavenging action of rising air bubbles from breaking waves
and rain drops (Grammatika and Zimmerman, 2001; Aller et al., 2005).

The sea-surface skin layer, in particular the microlayer, is subject to strong
diurnal variations in temperature, salinity, pH, and concentrations of
aldehydes and other organic molecules as a result of photochemical and
photo-biochemical interactions at the air-sea interface (Zhou and Mopper,
1997; Wootton et al., 2008).

Most of the heat and gas exchange between the ocean and the atmosphere of today
occurs from within the skin layer. The emitted infrared radiation from the sea originates 
from within the upper 100\thinspace \unit{\mu}m (Soloviev and Schl\"{u}ssel, 1994).

During the day, infrared (700--10\thinspace 000\thinspace nm), visible
(400--700\thinspace nm), and ultraviolet (290--400\thinspace nm) light is
absorbed at the sea-surface. Pure water has a low absorption coefficient for
visible and near UV light, as can be surmised from its transparency at these
wavelengths. However, the organic and inorganic material at the sea-surface
alters its optical properties such that an important part of
the total energy absorbed in the skin comes, in fact, from visible and UV light. Although,
to the authors knowledge, no measurements of the optical density of the
sea-surface skin layer have yet been published, an estimate can be made
given the fact that the sea-surface microlayer has a density of organic
material roughly 10$^{4}$ that of water slightly below (Grammatika and
Zimmerman, 2001), which is somewhat greater than the ratio of that for very %
\mbox{turbid} coastal waters to deep ocean water (Wommack and Colwell,
2000). Using, as a surrogate, the largest frequency dependent absorption
coefficient measured for turbid coastal waters of Bricaud et al.~(1981) and
assuming the solar spectrum at the Earth's surface for cloudless skies
(Gates, 1980), one can calculate that the organic material in the
sea-surface skin layer augments the absorption of energy in this layer by
about 13.3{\%} (9.1{\%} attributed to 290--400\thinspace nm UV absorption,
4.2{\%} attributed to 400--700\thinspace nm visible absorption) over that of pure
water, which absorbs predominantly in the infrared (Michaelian,
2010a). Under cloudy skies, or atmospheres of high water vapor content
(the probable condition on prebiotic Earth), infrared light is blocked and the effect of organic material
is much more important, increasing the energy absorption in the sea-surface
skin layer by a remarkable 400{\%} (200{\%} UV, 200{\%} visible)
(Michaelian, 2010a).

Absorption of infrared, visible, and UV light at the sea-surface today increases
the daytime temperatures at the skin surface by an average of  2.5\thinspace
K (up to 4.0\thinspace K) relative to the practically constant temperature
at a depth of 10\thinspace m (Schl\"{u}ssel et al., 1990). Night time
temperatures, on the other hand, are reduced on average by 0.5\thinspace K (up
to 0.8\thinspace K) due to evaporation and radiation. 
Cyanobacterial blooms have been shown to cause considerable
additional heating of the sea-surface (Kahru et al., 1993). The effect of
a nutrient enriched phytoplankton bloom on the energy exchange at the surface of a
lake has been quantified by Jones et al.~(2005). They found a 1.8\thinspace K
increase in the daytime surface temperature of the lake.

Very recent simulations by Gnanadesikan et al. (2010) using coupled climate models indicate 
that chlorophyll-dependent solar heating has a first-order impact on the spatiotemporal 
distribution of tropical cyclones. Not only the location of the cyclone, but both frequency 
and intensity are affected by chlorophyll surface concentration (Gnanadesikan et al., 2010).

Since the accumulation of organic material at the sea-surface is attributed
to surface tension, natural buoyancy and the scavenging effect of rising
bubbles, it is reasonable to assume that
an organically rich surface skin layer would also have existed in prebiotic
oceans. An estimate of the optical density of the skin layer in the
Archean can be made by assuming concentrations of $1.5\times 10^{-5}$%
\thinspace M/L for the nucleic acid bases, based on estimates by Miller
(1998) obtained from calculations by Stribling and Miller (1987) of
photochemical production rates of prebiotic organic molecules under
somewhat reducing conditions. Although these concentration estimates have
since been considered as optimistically large, the recent discovery of new
non-equilibrium abiogenic routes to these molecules (see below, and Powner
et al., 2009), and the possibility of oxidation inhibitors increasing yields 
substantially in a more neutral atmosphere (Cleaves et al., 2008), plus
the existence of an organically enriched sea-surface skin layer, may, in
fact, imply that they are underestimates. Using measured photon extinction
coefficients of the bases of around 13\thinspace 000\thinspace M$^{-1}$%
\thinspace L\thinspace cm$^{-1}$ at 260\thinspace nm (Chang, 2000), leads to
an absorption coefficient of 0.78\thinspace cm$^{-1}$. Using the Archean
solar spectrum at the Earth's surface as determined by Cnossen et al.~(2007),
the above determined UV absorption coefficient, and that of today for the
visible (probably an overestimate since pigments in the visible such as
chlorophyll did not exist in prebiotic oceans), leads to an increase in the
energy absorbed in the Archean sea-surface skin layer due to the nucleic
acid bases and other organics, over that of pure water, of about 19{\%} (11{\%} attributed to
200--300\thinspace nm UV absorption, 8{\%} attributed to 400--700\thinspace
nm visible absorption) for a cloudless day, and a remarkable 490{\%} (260{\%}
UV, 230{\%} visible) for an overcast day, or for an atmosphere with high water 
vapor content (Michaelian, 2010a).

Increased absorption of high energy photons at the
Archean sea-surface due to organic material would imply
lower albedo and an increase in the size of the water cycle. 
Today, about 46\% of the incoming solar radiation is absorbed at the 
surface of the Earth. About 53\% of this absorbed radiation is re-emitted into the 
atmosphere in the form of latent heat through evaporation and transpiration, 
while emitted long-wave
radiation accounts for 32\%, and direct conduction 15\%. Water vapor, on reaching the 
cloud tops condenses, releasing its latent heat of condensation in a black-body 
spectrum at -14$^\circ$C (Newell et al., 1974). Without organic material at the
sea-surface, more photons would be reflected, increasing the albedo, 
or penetrate deeper into the ocean, augmenting 
the bulk temperature of the oceans while leaving less heat available at the surface for
evaporation. More energy would then be emitted by the Earth as long-wave radiation
in a spectrum corresponding to the temperature of the ocean surface 
of about 18$^\circ$C, rather than at the lower temperature of the cloud tops, implying
less global entropy production. Terrestrial ecosystems would
also become less efficient at photon dissipation with less water in the global water cycle.
Incident angle and frequency dependent albedo for sunlight on ocean water of today has been
measured by Clarke et al.~(1970) and Jin et al. (2004). They find that organic material in ocean
water reduces the albedo at all incident angles and wavelengths and that
this effect increases significantly towards the shorter wavelengths of the
ultraviolet.

\section{Abiogenic synthesis of the molecules of life}

A central problem with all theories on the origin of life has been the
difficulty in demonstrating efficient abiogenic reaction pathways for
producing high yields of the primary molecules of life (Orgel, 2004). High
yields are important since the half-lives of these molecules are relatively
short at high temperature, on the order of hours for ribose, and years or
days for nucleic acid bases ($t_{1/2}$ for A and G $\approx $1\,yr; U $%
\approx $12\,yr; C $\approx $19\,days at 100\,$^\circ$C; Levy and Miller,
1998). Many of these molecules require chemical reactions which are
``uphill'', corresponding to overall positive changes in the Gibb's free
energy, while others have large activation barriers that require special
enzymes in order to proceed. Near equilibrium pathways to these molecules
have been found but do not lead to high yields.

Prigogine (1967), however, has shown that the yield of a product of a chemical
reaction can be increased enormously over its expected near-equilibrium
value by coupling the nominal reaction to other entropy producing
irreversible processes (Chang, 2000). Irreversible
processes can be coupled as long as the net production of entropy is
positive and Curie's principle is respected; macroscopic causes must
have equal or fewer elements of symmetry than the effects they produce. For
example, a chemical reaction (scalar) cannot give rise to a directed heat
flow (vector). These routes, under far from equilibrium conditions, have
scarcely been explored and may offer alternative pathways to efficient
abiogenisis (Wicken, 1978; Wicken, 1979; Pulselli et al., 2009).

For example, the second difficulty mentioned by Orgel (2004), the
polymerization of polynucleotide from mononucleotides is an 
endergonic reaction (positive free energy
change) which will not proceed spontaneously. However, this reaction can be
coupled with a second irreversible process, the absorption and dissipation
of a high energy photon, such that the overall reaction is exergonic
(negative free energy change). McReynolds et al. (1971) have shown that 
oligonucleotides can be produced by the action of UV light on an aqueous solution
of nucleoside phosphates. Such coupled photochemical reactions would have been
prevalent at the beginning of life given the greater amount of UV light
reaching the Earth's surface. Spontaneous polymerization of polynucleotide
under UV light is entropy driven since single strand RNA and DNA in
water is more efficient (rapid) in quenching the excitation energy of the
absorbed UV photon directly to the ground state through vibrational cooling
than are single (particularly pyrimidine) bases, which lose efficiency by
decaying to longer lived $^1n\pi ^\ast $ states about 40{\%} of the time
(Middleton et al., 2009).

Another important characteristic of ultraviolet light is that it can readily
destroy other organic molecules that have the potential for either
catalyzing the break-down of RNA and DNA, or for competing for reactants
needed for their synthesis. For example, Powner et al.~(2009) have found a
promising new route to pyrimidine ribonucleotide production, bypassing the
difficult production of ribose (the first problem mentioned by Orgel) and
free pyrimidine nucleobases, by employing UV light (254\thinspace nm) and a
heating and cooling cycle to enhance ribonucleotide synthesis over other
less endergonic products.

Other experimental data also support the
assertion that the prevalent conditions on Archean Earth -- intense UV light,
high temperature, and temperature cycling -- relieve the problem of the 
low yield of nucleotide synthesis. \mbox{Ponnamperuma et
al.~(1963)} have reported the detection of small amounts (0.01{\%}) of
adenosine when a 10$^{-3}$\thinspace M solution of adenine, ribose and
phosphate was irradiated with UV light. Folsome et al.~(1983) report on
anoxic UV photosynthesis of uracil, various sugars, including deoxyribose, and
amino acids. Kuzicheva and Simakov (1999) have shown that much larger yields
($\sim $4{\%}) of nucleotides can be synthesized by including temperature
cycling along with UV light. Their data were obtained by flying basic
compounds on board a spacecraft exposed to the UV and gamma environment of
space. The rotation of the spacecraft caused temperature cycling, an effect
to which they attribute higher yields than obtained in their laboratory
experiments with only UV light.

Finally, even though some reactions on the road to nucleotide synthesis may
be exergonic, their rates may be very low due to large activation barriers.
In this case, yields may be considerably increased by augmenting the
temperature (Ponnamperuma and Mack, 1965; de Graaf and Schwartz, 2005) 
or employing catalysts, such as enzymes. 

The thermodynamic forces of intense UV radiation and temperature cycling 
not only appear to alleviate the difficulty with yields
of the primary molecules, but are essential to a proposed mechanism of
ultraviolet and temperature assisted reproduction
of RNA/DNA without the need for enzymes.

\section{UV and temperature assisted RNA and DNA reproduction}

It is generally believed that RNA preceded DNA in life's
evolutionary history (RNA World hypothesis). This belief is based in part on
the fact that, because it is less stable, RNA exists more often in single
strand and shorter length segments than DNA, and can, therefore, fold in on
itself or pack together to form three dimensional structures akin to
proteins, which, under certain conditions, can catalyze chemical reactions.
For example, the active surfaces of ribosomes, the molecular machinery of
the cell where proteins are made, consist of RNA known as ribosomal RNA
(rRNA). An important catalytic activity of rRNA, which points to RNA as the
first molecule of life, is its demonstrated ability to catalyze peptide
bonds between amino acids to form proteins (Chang, 2000). On the other
hand, the lack of a hydroxol group on the ribose sugar of DNA allows it to
obtain its full three-dimensional conformation and to coil up to fit within
the nucleus of more modern eukaryote organisms, suggesting relevance to life
at a much later date. It is, therefore, reasonable to presume that RNA
preceded DNA. However, both molecules are produced with similar abiogenic
yields in vitro and both appear to have similar ultraviolet absorbing
and dissipating characteristics. Therefore, within the present framework, there is no 
overwhelming reason why DNA could not have replicated 
con-temporarily alongside RNA, performing the same function of
catalyzing the water cycle and entropy production through UV light absorption and dissipation.
The two molecules may have eventually formed a
symbiosis allowing new possibilities for more efficient reproduction and
correspondingly greater entropy production. 

The two naturally occurring
molecules RNA and DNA will, thus, be treated here on equal footing by
denoting both inclusive possibilities as ``RNA/DNA'', while acknowledging
that future data may favor one over the other as the first molecule of life
in the context of the proposed theory. Simpler synthetic molecules,
postulated as pre-RNA candidates, such as PNA, TNA and GNA (Egholm et al.,
1993), do not occur naturally and, therefore, probably have little to do
with photon absorption and dissipation in the biosphere.

At temperatures above 90\thinspace $^{\circ }$C (at one atmosphere and
pH~7), almost all of double strand RNA or DNA is denatured into
flexible single strands (Haggis, 1974). At lower temperatures, the amount of
denaturing depends on the relative proportion of G-C base pairs, 
the length of the strand, the pH of the solvent (very low and very high 
pH correlates with more denaturing; Williams et al., 2001), and the
salt concentration (higher salt concentration correlates with less
denaturing). RNA has generally lower denaturing temperature than similar
length DNA. Random nucleotide sequences and smaller length segments also
have lower denaturing temperature. For example, random RNA formed from equal
concentrations of A, G, C, and U has a melting temperature (defined as that
temperature at which half of the double strands are denatured) of
50\thinspace $^{\circ }$C, while calf thymus DNA has a melting temperature
of 87\thinspace $^{\circ }$C (Haggis, 1974). At the higher atmospheric
pressures thought to have existed at the beginning of life (up to twice the
present value) these denaturing temperatures may have been somewhat
higher.

At the high temperatures of the surface of the seas existing before the
beginnings of life on Earth, the nucleotides probably floated independently, 
unable to stack through Van der Waals and hydrophobic interactions,
or pair conjugate through hydrogen bonding, because of the large Brownian motion.
However, the Earth's surface gradually began to cool, and when the 
sea-surface temperature cooled to below that of the denaturing temperature of
RNA or DNA (relevant to the local prevailing ambient pressure, pH, and salinity) a
phenomenon, which may be called ``ultraviolet and temperature assisted
RNA/DNA reproduction'' (UVTAR), could have occurred.

One estimate has the surface temperature of the Earth descending below
100\thinspace $^{\circ }$C about 4.4\thinspace billion years ago (Schwartz
and Chang, 2002). Giant impacts, extending into the ``late lunar bombardment
era'' of ca.~3.9\thinspace Ga, may have periodically reset ocean
temperatures to above the boiling point (Zahnle et al., 2007). There is
geochemical evidence in the form of $^{18}$O/$^{16}$O ratios found in cherts
of the Barberton greenstone belt of South Africa indicating that the Earth's
surface temperature was $70\pm 15$\thinspace $^{\circ }$C during the
3.5--3.2\thinspace Ga era (Lowe and Tice, 2004). These surface temperatures,
existing at the beginnings of life (ca.~3.8\thinspace Ga),
are suggestively close to the denaturing temperatures of RNA/DNA.

During daylight hours, the water at the Archean sea-surface absorbed some
solar infrared light, and the aromatic bases of RNA/DNA and 
amino acids absorbed solar ultraviolet light, while other organic
molecules absorbed visible light. It is then probable that 
the sea-surface skin temperature
in the local neighborhood of the RNA/DNA would heat up beyond the
denaturing temperature and these would separate into single
strands by breaking the hydrogen bonds between conjugate base pairs.

RNA/DNA strongly absorb ultraviolet radiation at around 260\thinspace nm at
1~atmosphere pressure (Haggis, 1974; Chang, 2000) due to the $^{1}\pi \pi
^{*}$ electronic excitation of the bases (Voet et al., 1963; Callis, 1983).
The relaxation to the
ground state of UV excited DNA has been studied in detail by Middleton et
al.~(2009, and references therein). An ultra-fast, sub-picosecond, decay of
the $^{1}\pi \pi ^{*}$ excited state is observed for the unstacked bases in
single strand RNA/DNA through vibrational cooling to the ground state by
coupling to the high frequency modes of the water solvent  (Pecourt
et al., 2000, 2001). Water appears to
be the most efficient of many tested solvents (Middleton et al., 2009).
Such ultra-fast de-excitation does not appear to exist for stacked bases in
double strand RNA/DNA, which normally form long-lived, 100-picosecond,
exciton states. This may be partly due to the fact that hydrophobic
interactions exclude water from the interior of stacked double strand
RNA/DNA (Pecourt et al., 2000, 2001). 

It has
been suggested that these surprising characteristics are not fortuitous, but rather
remnants from earlier days when life was exposed to significantly higher
doses of UV radiation. The argument is that these characteristics would have been
favored by natural selection since such a highly efficient non-radiative
decay significantly lowers the rate of RNA/DNA damage through
photo-reactions, thereby reducing the need for frequent repair
(Crespo-Hern\'{a}ndez et al., 2004; Middleton et al., 2009). 

Sagan (1973) pointed out that the rapid UV photon dissipation characteristics of
nucleic acid bases would provide an important selective advantage to RNA and
DNA over other more easily synthesized organic molecules under the harsh UV
conditions of prebiotic Earth. Mulkidjanian et al.~(2003) have confirmed
this using simple Monte Carlo simulations. However, the
interpretation given here of these surprising absorption and relaxation
characteristics of RNA/DNA is more profound; apart from conferring stability
to these molecules under intense UV radiation, these characteristics confer
remarkable entropy producing potential to these molecules through the
efficient absorption of UV light and its rapid dissipation into heat.

As night came, with no light to absorb, the surface of the sea would
cool through evaporation, radiation, and conduction of heat to the atmosphere, 
to a temperature below which the single strands of RNA/DNA could
begin to act as templates and hydrogen bond through their bases with conjugate
nucleotides and oligonucleotides floating nearby. New,
complementary double-strand RNA/DNA would thus be formed at the sea
surface during the cool periods overnight. An alternative form of cooling of
the ocean surface may have been provided by hurricanes which are known to
have an important effect on reducing the surface temperatures of seas
(Manzello et al., 2007). Given the high sea-surface temperatures existing on
early Earth, and a cold upper atmosphere (Tian et al., 2005),
hurricanes would have been much more prevalent than at present.

As the Sun rose, about 7\thinspace h after setting (the rotation of the
Earth was more rapid 3.8\thinspace billion years ago) the sea-surface skin
layer would again heat up through the absorption of ultraviolet and visible 
light on the organic material, and the absorption on water of some infrared light
that could penetrate the clouds and water vapor in the atmosphere. Using the
Archean solar intensity spectrum at the Earth's surface as determined by Cnossen et al. (2007)
and the absorption coefficient for UV light between 200 nm and 300 nm of 
0.78 cm$^{-1}$ determined from estimated prebiotic concentrations of 
the nucleic acids (Miller, 1998; see section 3), 
while assuming similar absorption in the visible as that of today's oceans of $8.0\times10^{-3}$ ,
and that the diffuse downward and upward long-wave energy flux 
approximately cancel (Webster, 1994)  it can be estimated that diurnal temperature cycling
at the Archean sea-surface skin layer could have had an amplitude as large as 4\thinspace K (Michaelian, 2010a) . 
Schl\"{u}ssel et al. (1990) have measured a diurnal temperature variation at
the skin surface of today's oceans as large as 5\thinspace K. 

Direct absorption of a UV photon of
260\thinspace nm on RNA/DNA (which occurs preferentially on one or two of
the nucleic acid bases; Middleton et al., 2009) would leave 4.8\thinspace eV
of energy locally which, given the heat capacity of water, would be
sufficient energy to raise the temperature by an additional 3\thinspace K of
a local volume of water that could contain up to 50 base pairs (Michaelian,
2010a). Given that the full width of the denaturing curve for RNA/DNA is
between 4 and 10\thinspace K (depending on the G-C content) the sea-surface 
temperature in the neighborhood of the
segment which absorbed the UV photon would be raised again beyond the
denaturing temperature of RNA/DNA and the double strand would separate,
providing, in this way, a new generation of single strand RNA/DNA that could
serve as new template for complementary strand polymerization during the
subsequent cool period. Experimental evidence (Hagen et al., 1965, Roth and London, 1977) indicates 
that UV irradiation does indeed induce denaturation of DNA
held in water baths at a fixed temperature, and that the denaturing effect
of UV light increases as the temperature of the bath approaches DNA melting temperature.

A temperature assisted mechanism for RNA/DNA reproduction is not
hypothetical; the procedure of repetitive heating and cooling is a process
known as polymerase chain reaction (Mullis, 1990) that is used today in the
laboratory to amplify exponentially a particular DNA or RNA segment of
interest. The enzyme polymerase is used to speed up the polymerization of
nucleotides on the single strand templates during the low temperature period.

Ultraviolet and temperature assisted RNA/DNA reproduction would have been
enhanced by a number of natural phenomenon. First, single strand RNA/DNA
absorbs from 20{\%} to 40{\%} more ultraviolet light than does double 
strand. This effect, known as hypochromism (Bolton et
al., 1962; Chang, 2000), is related to the orientation of the electric
dipoles of the bases, stacked in fixed relation one above the other in the
double helix. On denaturation, the orientation of the dipoles is random and
the absorption intensity increases. Double
strand RNA/DNA is also less efficient (rapid) at transforming the electronic
excitation energy into heat than single strand randomly stacked DNA
(Middleton et al., 2009). Both these effects would provide positive feedback
for augmenting entropy production by stimulating denaturation under
solar UV light and by reducing the possibility of recombination of the
separated strands.

A second class of phenomena that could have enhanced UVTAR is the diurnal variation of the chemical
properties of the sea-surface microlayer. Both the pH and formaldehyde
concentration of the microlayer peak in late afternoon due, in part, to
causes predicted to be more relevant during the Archean; lower CO$_{2}$
dissolution in warmer water (Wootton et al., 2008) and increased UV
photochemical reaction rates (Zhou and Mopper, 1997), respectively. Both high 
pH and formaldehyde concentration promote lower DNA/RNA denaturing temperatures (Williams et
al., 2001; Traganos et al., 1975). Salinity also reaches a maximum at late
afternoon due to increased water evaporation in the microlayer (Zhang et
al., 2003), but this would produce a lesser effect opposing denaturation.

Since experimental determinations with the PCR technique give optimal
(specificity and rate) annealing temperatures of primers of about 5\thinspace K
 below DNA melting temperatures, these thermal and chemical
diurnal variations, as well as the longer denaturation and annealing
times allowed for by the UVTAR mechanism
(hours instead of minutes for PCR), suggest that an effective UV and temperature
assisted RNA/DNA replication mechanism could have been operating at the
Archean sea-surface. RNA/DNA at the beginning of life did 
not require enzymes for its
replication, reproduction was instead promoted by the day/night fluctuation
of the sea-surface skin temperature about the denaturing temperature
of RNA/DNA.

\section{Entropy production, information storage, and fidelity}

The link between entropy and information has been made by Shannon and Weaver
(1949). However, information only has thermodynamic relevance in the context
of its ability to catalyze irreversible processes. In a non-equilibrium
environment with thermodynamic forces over the system, information is thus
more correctly associated with entropy production.

As the seas began to cool further, competition for free nucleotides
would imply that those RNA/DNA segments which, through particular base 
sequences along their length, had lower denaturing
temperatures or could absorb more ultraviolet light and transfer this
energy more efficiently to the molecular degrees of freedom of the 
surrounding sea-surface water, would be those
favored for reproduction because of a higher denaturing probability. For example, 
favored segments would be those
with more adenine-uracil (A-U) pairs for RNA, or adenine-thymine (A-T) pairs
for DNA, as opposed to those with more guanine-cytosine (G-C) pairs, since
the later have higher denaturing temperatures due to stronger van der Waals
interactions between neighboring G-C pairs and because the later are
joined by three hydrogen bonds while A-T pairs have only two. For example, a
DNA segment containing 30{\%} of G-C pairs has a denaturing temperature of
82\,$^\circ$C while that containing 60{\%}  denatures at 96\,$%
^\circ$C (Chang, 2000).

This might lead to the conclusion that, as the sea temperature cooled to
below the denaturing temperature of RNA/DNA, the composition of the organic
soup would consist of mainly long double-strand RNA/DNA containing a
high percentage of G-C
pairs that could no longer replicate, and many more very short-strand
RNA/DNA containing a preference for A-U or A-T pairs that could continue
replicating, and thereby increase their proportionate representation. 
However, there is an important possibility which may have arisen to counter 
this bias. If, by chance, a codon of the longer RNA/DNA strands coded 
for a simple enzyme that could help it denature
at a lower temperature, then these longer strand RNA/DNA would retain their
replicating ability and thus their selective advantage even in an
ever colder sea. For example, the easily abiogenically synthesized aspartic
amino acid (Asp) is a metabolite of the urea cycle that can produce urea
from ammonia. Urea (and also formamide and formaldehyde) in the presence of magnesium ions
can reduce DNA melting temperatures by about 0.6\,$^\circ$C for
every 1{\%} increase of the denaturing substance (Jungmann et al., 2008).
The thermodynamic advantage of maintaining, and even enhancing, the entropy
production in ever colder seas could thus have been the origin of the
information content and the reproduction fidelity of RNA/DNA.

The association of a RNA/DNA molecule with an enzyme, or protein, is
still in very common existence today, known as a virus. Poteins absorb strongly at
280\thinspace nm due to the aromatic ring of their constituent amino acids
tyrosine, tryptophan and phenylalanine (Haggis, 1974; Chang, 2000). 
They could thus have first served as antenna
molecules by providing a larger cross section for UV absorption, and thereby
more local heating, favoring denaturing and thus still greater entropy production.

\section{Evolution}

With greater cooling of the oceans came greater selection pressure for
longer RNA/DNA segments that could code for still more complex enzymes that
could facilitate denaturing at still cooler temperatures. Just as urea and
formamide require the presence of magnesium ions to lower the denaturing
temperature of DNA, a denaturing enzyme existing today, helicase, requires a
magnesium ion for coupling adenosine triphosphate (ATP) hydrolysis to
nucleic acid unwinding (Frick et al., 2007). Proteins and other molecules 
containing a magnesium ion would have had a natural affinity to RNA/DNA 
because of their ionic attraction to the negatively charged phosphate groups of 
the backbone.

Magnesium ions are also an important component of the pigment chlorophyll.
The most readily abiogenically synthesized amino acid glycine reacts with
succinyl-CoA from the citric acid cycle to form a porphyrin which, when
coordinated with a magnesium ion, forms chlorophyll.

Another possible product of information for promoting UV and temperature
assisted RNA/DNA reproduction as the seas cooled is the coding for sequences
of RNA that promote self-splicing (eg. self-splicing introns), or in the
case of DNA, for a primitive topoisomerase enzyme which can break lengths of
DNA into shorter parts for a transient time, effectively giving a large
RNA/DNA the lower denaturing temperature of the smaller length segments. The
catalytic residue of the topoisomerase enzyme is based on the amino acid
tyrosine (Tyr). This amino acid is also used in the photosystem II of
chloroplasts, acting as the electron donor of the oxidized chlorophyll.
Tyrosine, because of its aromatic ring, absorbs strongly at 280\,nm
(Chang, 2000) and, like RNA/DNA in water, has demonstrated chemical
stability under high doses of UV radiation (Barbiera et al., 2002),
suggesting that its initial association with RNA/DNA may have been as a
robust antenna type of photon absorber to augment the local water
temperature sufficiently for denaturation.

It is thus possible that RNA/DNA segments containing the codons specifying
for one or more of these amino acid cofactors, acting as primitive
denaturing enzymes, gradually mutated into one specifying for a primitive
Chlorophyll molecule. With the new molecule Chlorophyll allied with RNA/DNA,
the efficacy of the light absorption and dissipation process would be still
further enhanced as the skies cleared of organic haze, aldehydes,
cyano-molecules, sulfur dioxide and sulfuric acid and water clouds,
permitting more visible radiation to penetrate. Chlorophyll absorbs strongly 
at 430\,nm, just where water is most transparent.

RNA/DNA segments that mutated to code for enzymes that could help it
capture, join, and polymerize with individual free floating nucleotides
would then find even greater selective advantage in still colder seas. Due 
to reduced Brownian motion, an accidental, and correctly aligned,
meeting of a RNA/DNA single strand with a free floating nucleotide or oligonucleotide would
become ever more improbable. It is relevant again here that the nucleic acid
polymerase of today contain Zn$^{++}$ and Mg$^{++}$ ions as cofactors for
their enzymatic activity (Rauchfuss, 2008). As the rain of the nucleic acid
bases and other organic molecules from the sky began to dry up, RNA/DNA
segments that coded for enzymes that could not only capture but also
synthesize the bases from more primitive but more prevalent organic
molecules such as hydrogen cyanide (Matthews, 2004), would be increasingly
more favorably selected.

Since ATP is synthesized in the chloroplasts of plants of today by a process
known as photophosphorylation, there may have existed a more primitive
direct photochemical rout to its synthesis involving UV light and heat. For
example, Muller (2005) suggests that thermosynthesis of ATP may be possible
through temperature cycling in hydrothermal vents. However, UV light
absorption and dissipation by the nucleic acid bases during daylight hours
and cooling of the surface at night by evaporation and radiation, 
might have provided a similar heat
engine for the abiogenic synthesis of ATP. For example, Kuzicheva and Simakov (1999) have
measured significant yields of 5'AMP from nucleosides and inorganic
phosphates due to the action of UV light and temperature cycling on
spacecraft experiments.

It is a remarkable fact that the protein bacteriorhodopsin, that promotes
ATP production in archea by acting as a proton pump through the absorption
of a photon at 568 nm in the visible, also works perfectly well by absorbing
at 280 nm in the ultraviolet (Kalisky et al., 1981). The UV photon energy is
absorbed on the aromatic amino acids tyrosine and tryptophan and the energy
transmitted to the chromophore. It may thus have been that ATP originally
obtained its free energy for formation directly from UV light and that life
found a means to harvest this stored free energy for endergonic reactions
that required a softer mode (or delayed mode) of energy transfer that
could not be provided by direct UV or visible photons. 

Thus may have begun the history of the evolution of amino acids, proteins,
and ATP with RNA/DNA; first as simple catalysts to help it denature at
colder sea temperatures, as complements that aided in the absorption of UV
and visible light, and finally as more active catalysts in attracting
cyano-molecules, synthesizing nucleotides, and polymerizing RNA/DNA, as
well as providing an active unwinding mechanism as RNA/DNA grew in length and sea
temperatures cooled. The increasing competition between the reproducing
RNA/DNA segments for organic molecules produced in the Earth's early
atmosphere, or delivered from outer space by the comets and meteorites, and
the importance of this reproduction to the entropy production of the Earth,
could thus have constituted the first steps of evolution through natural
selection.

Since the life induced changes in the composition of the Earth's atmosphere,
the rain of organic molecules from the sky has now ceased. Viruses, which
may thus have been the remnants of the very beginnings of life, and which
were able to obtain their component molecules from abiogenisis and reproduce
through the UVTAR mechanism, have now evolved to parasitize the complex nucleic acid
production of existing life within the protected environment of the cell
(Smith, 1965). Their associated proteins may have evolved from simple
denaturing enzymes to vehicles that facilitate entry across cellular
membranes. Their principal function appears to have also changed from being
dissipators of ultraviolet light to agents that cause mortality through lysing
in bacteria and higher organisms, allowing their nutrients to be recycled
into photosynthetic life. It has been suggested that viruses have co-evolved
with primordial cells, stimulating them to produce rigid walls, which caused
genetic privacy, allowing vertical evolution (Jalasvuori and Bamford, 2008).
Vertical evolution has also augmented entropy production on Earth through
the water cycle by producing higher mobile life forms which by transporting
nutrients allowed the photon dissipating molecules to spread into regions
initially inhospitable to their reproduction (Michaelian, 2009).

Viruses have recently been found to be much more prevalent in sea water than
suspected; being the major component by number of the organic material in
ocean water (Wommack and Colwell, 2000). These so called ``virioplankton''
are found in concentrations of 10$^{4}$ to 10$^{8}$ per ml of ocean water,
usually about one to two orders of magnitude more common than the bacterial
phytoplankton that they parasitize. They are also found in highest abundance
within the first 20\,%
\unit{\mu}m of still lake water (Wommack and Colwell, 2000). Viral DNA
appears to account for only approximately 20{\%} of the dissolved DNA found
in ocean water. The rest appears to be soluble DNA of roughly 500 base pairs
in size (viral DNA has greater than 20 thousand base pairs) of still unknown
origin but probably the result of virus lysing bacteria (Wommack and
Colwell, 2000). Bacterial lysing by viruses 
augments the production of cyanobacteria and other phytoplankton
by cycling through necessary nutrients such as phosphor and nitrogen
(Wommack and Colwell, 2000; and references therein). This floating organic
material, in effect, increases the absorption and dissipation of sunlight at
the surface of lakes and oceans today (Jones et al., 2005; Michaelian,
2010a).

\section{Comparison with prevailing scenarios for the origin of life}

Traditional views of the origin of life see it as an extraordinary event, 
persisting for a kind of inherent self-indulgence. The
present theory sees the origin of life as 
a thermodynamic imperative; a new irreversible process, arising once 
environmental conditions became appropriate, that coupled to existing abiotic irreversible
processes to augment the global entropy production 
of the Earth in its interaction with its  solar environment. Although prevailing 
theories recognize the necessity of a free energy source for the metabolism of 
self-perpetuation, such as that derived 
from chemical potential or thermal gradients of a hydrothermal vent (W\"achterh\"auser, 2006),
they fail to acknowledge a greater thermodynamic function of life beyond
self-perpetuation; dissipation of the solar photon flux through life's coupling to the water cycle. 

The solar photon flux would have been the most intense and extensive 
free energy source available during the Archean. The present theory suggests how the 
visible photon dissipation by life of today may be directly linked with UV photon dissipation at 
life's beginnings in the Archean. Chemical and thermal 
gradient sources of free energy would have been limited in size and
extent and there would remain the difficult problem of rapidly evolving photon 
dissipation, appearing shortly after the origin of life, from chemical dissipation.  

On addressing evolution, prevailing theories are subject to the same criticism of 
circularity of argument that inflicts Darwinian theory. What is natural selection selecting? 
The present theory suggests that natural selection selects biotic-abiotic 
coupled systems leading to ever greater global entropy production, in accordance with
the Onsager principle and empirical evidence.

The present theory shares with hydrothermal vent proposals congruency 
with evidence supporting a thermophilic origin of life (Schwartzman and Lineweaver, 2004). 
Any irrefutable evidence indicating that the surface of the Earth was cold at the 
time of the origin of life would render the theory inviable. The present theory suggests
that life as we know it may be less universal than prevailing theories might indicate,
since, besides the usual requirements of liquid water, chemical constituents, and a free
energy source, the theory also imposes stringent requirements on the boundary, initial, and evolving 
external conditions; an atmospheric window allowing  a high UV flux to reach the surface, 
temperatures descending gradually below the melting temperature of RNA/DNA, 
temperature cycling, and a water cycle.

The present theory avoids the difficulty of the RNA World hypothesis that has made
it most vulnerable to criticism; that of requiring  a priori 
sufficient RNA information content and reproductive fidelity (Shapiro, 2007), 
employing instead the particular environmental conditions of Archean Earth. 
A temperature cycling mechanism for amplification of 
RNA and DNA indeed has been shown to function in the laboratory (polymerase chain reaction).

In contrast to prevailing theories, the present theory does not require the 
unlikely discovery of an abiotic mechanism that produced an initial high 
enrichment of chiral enantiomers to explain the homochirality of life today. 
Instead, the present theory argues that homochirality arose gradually over time due to a 
small asymmetry in the environmental conditions that promote the UVTAR mechanism.  
In particular, if indeed life emerged when the
sea-surface temperature had cooled to below the denaturing temperature of
RNA/DNA, then, since the sea-surface temperature would be greatest in the
late afternoon, the absorption of the slightly right-handed circularly polarized 
light of the afternoon (Angel et al., 1972), could have contributed to the abundance of
RNA/DNA with D-enantiomer nucleotides. Double strands containing 
predominantly L-enantiomer nucleotides would absorb less well the
right-handed circularly polarized light, and thus could not raise local water 
temperatures as often for denaturation, thereby curtailing their replication and 
evolution (Michaelian, 2010b).

\conclusions[Discussion and conclusions]

Competition for organic
molecules in itself could not have led to evolution through natural
selection, or even to simple reproduction. This has been made clear from the
numerous experiments of Orgel and others which have failed to create self
replicating systems in the laboratory (Orgel, 1994, 2004).  Ignoring the 
entropy producing function of life is, in fact, the basis of the 
tautology in Darwin's theory of evolution through natural selection. 
As Boltzmann hinted 150 years ago, the vital force of life and evolution is
derived from photon dissipation, i.e.~through entropy production. Greater
numbers of RNA/DNA in the Archean absorbed more sunlight and catalyzed the early
Earth water cycle, besides driving ocean and wind currents. Reproduction and
evolution were thus synonymous with increases in the entropy production of
the coupled biotic and abiotic biosphere. Naturally selected mutations of
the RNA/DNA-protein complexes, and later that of complex animals and
ecosystems, would be those allowing for ever greater increases in absorption
of high energy photons and greater efficiency at converting these into heat.

The non-equilibrium thermodynamics of the abiogenisis of the primary molecules
and their polymerization has hitherto not been duly
considered in theories addressing the origin of life. High temperatures,
temperature cycling, and UV light have been demonstrated to be useful in
augmenting the abiogenic yields of the nucleotides and the
polynucleotides.

The problem of RNA/DNA replication without the use of enzymes has been
addressed through an ultraviolet and temperature assisted mechanism
involving cycling of the primitive sea-surface skin temperature around the
denaturing temperature of RNA/DNA and the remarkable ability of these
molecules to absorb and dissipate rapidly into heat the intense UV light that
penetrated the primitive Earth's atmosphere. The formation and
replication of RNA/DNA would be thermodynamically favored
because of the overall increase in entropy production that these molecules
afforded to coupled biotic-abiotic irreversible process occurring in the
biosphere, in particular, to the water cycle.

The origin of information content and replication fidelity of RNA/DNA
could be conceived of within the framework of the proposed theory if particular
polynucleotide sequences coded for enzymes that facilitated
denaturation at colder sea temperatures, or for light harvesting antenna
molecules, leading to a differential replication success and entropy producing potential of
different sequences. This may have been the beginnings of evolution through
natural selection.

The problem discussed by Orgel (2004) of the racemic product of
nucleotides frustrating the copying of stable polynucleotides has also been
addressed by the present theory. The slightly right-handed circularly polarized light 
of the late afternoon would have led to homochirality
before surface temperatures had cooled to a point at which enzymes and, 
therefore, reproductive fidelity, were necessary. This
completes the consideration from within the proposed framework of the
major problems concerning the origin of life as reviewed by Orgel (2004).

A first step in probing the veracity of the proposed theory would be to test
experimentally if polymerase chain reaction (PCR) could be carried out by
substituting the heat cycling thermostat with UV light cycling, with the
thermal bath held constant at a few degrees below the melting
temperature of the short ($<$50\,bp) strand RNA/DNA segment, or with slight
temperature cycling ($\sim $5\,$^\circ$C) representing day and night
temperature fluctuations of the Archean sea-surface.

The origin of life and beginnings of evolution, as depicted by this theory
has the general feature of an auto-catalytic cycle involving a strong
coupling between biotic and abiotic processes, driven by the goal oriented
and universal process of increasing the entropy production of Earth in its
interaction with its solar environment. This great auto-catalytic cycle
involving life and abiotic entropy producing processes remains to this day,
and appears to be evolving towards still greater efficiency at producing
entropy. Since the appearance of chlorophyll, new pigments
capable of capturing ever more of the Sun's spectrum have been incorporated
into the photosynthesizing systems of today's plant and
bacterial life. Examples are the carotenoids in green plants, the
phycobilins in phytoplankton, and the recently discovered mycosporine-like
amino acids (MAA's) in phytoplankton which absorb across the ultraviolet
(Whitehead and Hedges, 2002). Most of these pigments are known not to have a
direct role in photosynthesis. Furthermore, a number of complex mechanisms exist in plants
today to dissipate into heat photons absorbed in excess. These pigments and
mechanisms have hitherto
been considered merely as ``safety valves'' for photosynthesis (Niyogi,
2000). Alternatively, they may now be explained on
thermodynamic grounds through their importance to photon dissipation and
evaporation of water (Michaelian, 2009). 

The net effect of the origin and
evolution of life on Earth has been to gradually increase the Earth's
entropy producing potential, or, in other words, to reduce the Earth's
albedo and effective temperature at which it emits infrared radiation,
making it ever more a blackbody of lower temperature.

\begin{acknowledgements}
 The financial assistance of DGAPA-UNAM, 
grant numbers~IN118206 and~IN112809 is greatly appreciated.
\end{acknowledgements}

\end{document}